\documentclass[journal]{IEEEtran}

\ifCLASSINFOpdf
\else
   \usepackage[dvips]{graphicx}
\fi
\usepackage{url}
\usepackage{graphicx}
\usepackage{enumerate}
\usepackage{breqn}
\usepackage{algorithm,cite}
\usepackage[noend]{algpseudocode}
\usepackage{multirow}
\usepackage{soul,color}
\usepackage{balance}
\usepackage{amsmath,amsthm,amssymb}
\usepackage{hyperref}
\usepackage{enumitem}

\begin{document}

\title{RSD-GAN: Regularized Sobolev Defense GAN Against Speech-to-Text Adversarial Attacks}

\author{Mohammad Esmaeilpour,~\IEEEmembership{Senior Member,~IEEE,}
Nourhene Chaalia,~\IEEEmembership{Senior Member,~IEEE,}
and Patrick~Cardinal,~\IEEEmembership{Senior Member,~IEEE}
\thanks{M. Esmaeilpour, and P. Cardinal are with \textbf{\'{E}cole de Technologie Sup\'{e}rieure (\'{E}TS)}, Universit\'{e} du Qu\'{e}bec, Montr\'{e}al, Qu\'{e}bec, Canada. M. Esmaeilpour and N. Chaalia are with the Fédération des Caisses \textbf{Desjardins} du Québec. Contact: mohammad.esmaeilpour@desjardins.com.

This work was funded by Fédération des Caisses \textbf{Desjardins} du Québec, \textbf{IVADO} Institution, and \textbf{Mitacs} accelerate program with agreement number IT27051. {Supplementary material and source codes are available at \href{https://github.com/EsmaeilpourMohammad/RSD-GAN.git.}{this github-Repo.}}}}

\markboth{Accepted for publication in IEEE Signal Processing Letters, September 2022}
{Shell \MakeLowercase{\textit{et al.}}: Bare Demo of IEEEtran.cls for IEEE Journals}
\maketitle

\begin{abstract}
This paper introduces a new synthesis-based defense algorithm for counteracting with a varieties of adversarial attacks developed for challenging the performance of the cutting-edge speech-to-text transcription systems. Our algorithm implements a Sobolev-based GAN and proposes a novel regularizer for effectively controlling over the functionality of the entire generative model, particularly the discriminator network during training. Our achieved results upon carrying out numerous experiments on the victim DeepSpeech, Kaldi, and Lingvo speech transcription systems corroborate the remarkable performance of our defense approach against a comprehensive range of targeted and non-targeted adversarial attacks.               
\end{abstract}


\IEEEpeerreviewmaketitle

\section{Introduction}
\label{sec:intro}
Over the last decade, a remarkable progress has been made in developing real-time speech-to-text transcription systems particularly after the proliferation of deep learning algorithms. Nowadays, such systems employ variants of complex configurations such as long short-term memory, residual, and transformer architectures \cite{vila2018end} for improving transcription accuracy under severe noisy environments. DeepSpeech \cite{MozillaImplementation}, Kaldi \cite{povey2011kaldi}, and Lingvo \cite{shen2019lingvo} are among those advanced systems to name a few. However, it has been demonstrated that they are extremely vulnerable against carefully crafted adversarial signals which carry imperceptible perturbations ($\delta$) that redirect the recognition models towards transcribing incorrect phrases \cite{schonherr2020imperio}. Technically, an adversarial signal $\vec{x}_{\mathrm{adv}}$ can be defined as \cite{carlini2018audio}:
\begin{equation}
\vec{x}_{\mathrm{adv}} \leftarrow  \vec{x}_{\mathrm{org}} + \delta
\end{equation}
\noindent where $\vec{x}_{\mathrm{org}}$ denotes the original signal. For achieving an optimal value for $\delta$, often a costly optimization formulation is required since the difference between $\vec{x}_{\mathrm{adv}}$ and $\vec{x}_{\mathrm{org}}$ should be inaudible. Toward satisfying this condition, many optimization algorithms have been developed thus far. However, the majority of them implement a sort of convex formulation most likely inspired by Carlini {\it et al.}~\cite{carlini2018audio} as the following (C\&W attack).
\begin{dmath}
    {\min_{\delta} \left \| \delta \right \|_{F} + \sum_{i}c_{i}\mathcal{L}(\vec{x}_{\mathrm{adv}},\hat{\mathbf{y}}_{i})} \\ {\mathrm{s.t.} \quad l_{\text{dB}}(\vec{x}_{\mathrm{adv}}) < \epsilon \quad \mathrm{and} \quad \hat{\mathbf{y}}_{i} \neq \mathbf{y}_{i}}
    \label{eq:generalformulation}
\end{dmath}
\noindent where $c_{i}$ and $\mathcal{L}(\cdot)$ refer to adjusting coefficient (scalar) and the loss function for the victim transcription model, respectively. Additionally, $\mathbf{y}_{i}$ and $\hat{\mathbf{y}}_{i}$ denote the original and adversarial phrases in which the latter should be defined by adversary. Furthermore, $l_{\text{dB}}(\cdot)$ is a dB-scale loudness metric for measuring the distortion of $\vec{x}_{\mathrm{adv}}$ relative to $\vec{x}_{\mathrm{org}}$. Finally, $\epsilon$ is a subjective threshold for the adversarial perturbation which should be carefully tuned according to the properties of the transcription model, environment, and the characteristics of the original input signals (e.g., the number of channels).

Unfortunately, adversarial signals pose major security concerns for speech transcription systems in two aspects. Firstly, they can significantly reduce the recognition accuracy and generalizability of such systems \cite{wang2020towards}. Secondly, they are transferable from a model to another \cite{papernot2018characterizing}. For instance, an adversarial signal which is crafted for Kaldi, most likely fools other systems such as DeepSpeech and Lingo \cite{esmaeilpour2020sound,esmaeilpour2022multi}. Motivated by these two raised concerns, we develop a novel defense algorithm for counteracting with varieties of white and black-box adversarial attacks. Technically, our approach belongs to the synthesis-based defense category using generative adversarial network (GAN) \cite{goodfellow2014generative}. Briefly, this category includes defense algorithms which synthesize a new signal acoustically very similar to the given input speech ($\vec{x}_{\mathrm{in}}$) aiming at bypassing the potential adversarial perturbation on $\vec{x}_{\mathrm{in}}$. This is one of the reliable approaches for protecting recognition models \cite{samangouei2018defensegan}. In summary we make the following contributions in this paper: (i) Developing a novel regularizer for the integral probability metric of the defense GAN in the restricted Sobolev space \cite{van1983matrix,mroueh2017sobolev}; (ii) Incorporating the room-impulse-response (RIR) effect into the regularization formulation to enhance the quality of the synthetic signals; (iii) Optimizing the functionality of the discriminator network in a GAN setup for significantly reducing the number of required gradient computations. The rest of our paper is organized as the following. Section~\ref{sec:backg} provides a brief review over the state-of-the-art attack and defense algorithms. In Section~\ref{sec:proposedmethod}, we introduce our defense approach and explain its functionality. Finally, we summarize all the experimental results in Section~\ref{sec:exp}.

\section{Background: Attack and Defense}
\label{sec:backg}
Over the last years, many variants have been introduced for Eq.~\ref{eq:generalformulation} both in the context of targeted and non-targeted as well as white and black-box adversarial attacks (more information about these taxonomies are available at \cite{akhtar2018threat}). For instance, Yakura {\it et al.}~\cite{yakura2018robust} demonstrated that adversarial perturbation fades away upon replaying $\vec{x}_{\mathrm{adv}}$ over the air. For tackling this issue, they proposed an expectation over transformation (EOT) operation and slightly updated Eq.~\ref{eq:generalformulation} as follows:
\begin{equation}
    \min_{\delta} \mathbb{E}_{t\in \tau, \omega\sim \mathcal{N}(0,\sigma^{2})}\left [ \mathcal{L} ( \mathrm{mfcc}(\vec{x}_{\mathrm{adv}}),\hat{\mathbf{y}}_{i})+\alpha_{t}\left \| \delta \right \| \right ]
    \label{eq:yakura}
\end{equation}
\noindent where $t$ and $\tau$ denote an independent variable and EOT filter set, respectively. Additionally, $\omega$ represents the white Gaussian noise operator driven from the RIR simulation \cite{habets2006room} where $\vec{x}_{\mathrm{org}}$ is recorded. Such an operator is essential in optimizing for $\delta$ since it contributes to smoothly (i.e., in an imperceivable manner) contaminate the original input signal with the adversarial perturbation. In Eq.~\ref{eq:yakura}, $\mathrm{mfcc}$ is the Mel-frequency cepstral coefficient transform \cite{yakura2018robust,davis1980comparison} for converting $\vec{x}_{\mathrm{adv}}$ into a feature representation. Finally, $\alpha_{k}$ refers to the scaling parameter for adjusting the magnitude of $\delta$ in conjunction with the probability density function of $\tau$. 

A more simplified yet effective version for Eq.~\ref{eq:yakura} is called the robust attack and it has been introduced by Qin {\it et al.}~\cite{qin2019imperceptible}. This attack replaces the costly $\omega$ operation with a mask loss function for encoding $\delta$ beyond the human audible range. However, it does not incorporate the properties of microphone-speaker (SM) settings during its optimization procedure. It has been experimentally demonstrated that such settings effectively enhance the robustness of $\delta$ after consecutive playbacks over the air \cite{schonherr2020imperio}. Presumably, based on the concept of SM simulation, the Imperio \cite{schonherr2020imperio} and Metamorph \cite{chen2020metamorph} attack algorithms have been developed. The latter attack also simulates the channel impulse response (CIR) for achieving a less likely detectable adversarial perturbation upon running potential defense algorithms. However, CIR implementation is computationally expensive and requires running numerous exploratory experiments to tune its hyperparameters. To tackle this issue, Esmaeilpour {\it et al.}~\cite{esmaeilpour2022towards} introduced a novel attack formulation based on Cram\'{e}r integral probability metric (CIPM). Furthermore, their attack does not exploit any EOT operation during optimizing for $\delta$ and it converges relatively faster than other aforementioned algorithms.

All the above-mentioned adversarial attack approaches have been widely experimented and analyzed on the DeepSpeech (Mozilla's implementation), Lingvo, and Kaldi. Unfortunately, it has been evidently revealed that such cutting-edge systems are extremely vulnerable against almost all variants of speech attacks \cite{hu2019adversarial, esmaeilpour2021towards, carlini2018audio}. In response to this critical concern, a number of defense algorithms have been developed which we briefly review them in the following.

Multi-rate compression algorithm (MRCA) \cite{das2018adagio} is one of the baseline defense approaches which has been extensively used for benchmarking the resiliency of adversarial attacks under different noisy environments. The intuition behind this approach is straightforward. In a nutshell, MRCA compresses the input signal through high-frequency components modulation aiming at fading out the potential adversarial perturbation on $\vec{x}_{\mathrm{adv}}$. Although MRCA is computationally very efficient, it fails to detect adversarial signals which have been carefully optimized through CIR-based algorithms \cite{esmaeilpour2020class}. For tackling this challenge, two affirmative resolutions have been devised: (i) augmenting the transcription models with some simulated psychoacoustic filters which is known as the Dompteur defense \cite{eisenhofer2021dompteur}, and (ii) developing an autoencoding GAN (A-GAN) \cite{latif2018adversarial} for synthesizing new $\mathrm{mfcc}$ features for every input signal ($\vec{x}_{\mathrm{in}}$). Relatively, the latter defense approach enables reconstructing higher quality speech samples without recovering potential $\delta$. However, it has been shown that A-GAN suffers from extreme instability and mode collapse issues particularly for long speech signals ($> 5.6$ seconds) \cite{esmaeilpour2021towards}.

Towards developing stable GANs for effectively protecting speech transcription systems, a varieties of approaches have been presented heretofore. For instance, the class-conditional defense GAN (CCD-GAN) \cite{esmaeilpour2021cyclic} proposes a novel regularization technique as:
\begin{equation}
    \arg \min_{\mathbf{z}_{i}} \left \| \gamma\left [ \mathcal{G}(\mathbf{z}_{i}),\mathbf{x}_{\mathrm{org}} \right ] \right \|_{2}^{2} \quad \mathrm{and} \quad \mathbf{z}_{i} \in \mathbb{R}^{d_{z}}
    \label{eq:chorldal}
\end{equation}
\noindent where $d_{z}$ refers to the dimension of $\mathbf{z}_{i}$ and $\mathbf{x}_{\mathrm{org}}$ denotes the spectrogram (frequency components plot) \cite{mcfee2015librosa} associated with $\vec{x}_{\mathrm{org}}$. Additionally, $\mathcal{G}(\cdot)$ indicates the generator network in the GAN configuration and $\gamma\left [ \cdot \right ]$ refers to the chordal distance adjustment operator \cite{esmaeilpour2020detection,van1983matrix}. Eq.~\ref{eq:chorldal} iteratively maps $\mathcal{G}(\mathbf{z}_{i})$ onto the subspace of $\mathbf{x}_{\mathrm{org}}$ for ensuring that the projected signal does not carry any malicious perturbation (i.e., $\delta$).

In order to make Eq.~\ref{eq:chorldal} more compatible with other defense GAN platforms such as the complex cycle-consistent configurations, a convex quadratic regularizer has been introduced as \cite{esmaeilpour2021cyclic}:
\begin{equation}
    \arg \min_{\mathbf{z}_{2,i}} \left \| \mathcal{G}_{1}(\mathbf{z}_{1,i}^{c})-\mathcal{G}_{2}(\mathbf{z}_{2,i}) \right \|_{2}^{2} \quad \mathrm{and} \quad \mathbf{z}_{1,i}, \mathbf{z}_{2,i} \in \mathbb{R}^{d_{z}}
    \label{eq:cdgan}
\end{equation}
\noindent where $\mathcal{G}_{1}(\cdot)$ and $\mathcal{G}_{2}(\cdot)$ represent two generator networks bundled together sequentially. This regularization technique has been exploited in the cyclic defense GAN (CD-GAN) \cite{esmaeilpour2021cyclic} platform and the achieved results corroborate that not only Eq.~\ref{eq:cdgan} improves the stability of the entire generative model, but also it effectively enhances the chance of synthesizing $\delta$-free speech signals.

Designing a GAN architecture with multiple discriminator networks is another efficacious approach for improving the stability of $\mathcal{G}_{i}(\cdot)$. On one hand, this design contributes to provide more informative gradients to the generators and consequently enhances the performance of the defense algorithm in runtime \cite{hosseini2018multi,esmaeilpour2021towards}. On the other hand, such a design adds computational overhead to the training procedure of the generative model. Presumably, one potential resolution is defining the regularization term (e.g. Eq.~\ref{eq:cdgan}) in a non-Cartesian vector space \cite{freeman1970efficient,van1983matrix}. Following this perspective, Esmaeilpour {\it et al.}~\cite{esmaeilpour2022multi} developed a multi-discriminator GAN in the restricted Sobolev space \cite{brezis2010functional}. This defense algorithm is known as SD-GAN and it has been successfully evaluated for varieties of targeted and non-targeted adversarial attacks. However, it has been experimentally demonstrated that SD-GAN negatively affects the quality of the synthesized speech signals. In response, we develop a novel GAN-based defense algorithm which makes a better trade-off between protecting the transcription system against adversarial attacks and preserving the signal quality.

\section{Regularized Sobolev Defense GAN: RSD-GAN}
\label{sec:proposedmethod}
Typically, a GAN configuration consists of two deep neural networks, namely a generator and a discriminator denoted by $\mathcal{G}(\cdot)$ and $\mathcal{D}(\cdot)$, respectively \cite{goodfellow2014generative}. The first network enquires a random vector $\mathbf{z}_{i}$ and synthesizes a new signal $\vec{x}_{\mathrm{syn},i}$ as \cite{goodfellow2014generative}:
\begin{equation}
    \mathcal{G}(\mathbf{z}_{i}) \mapsto \vec{x}_{\mathrm{syn},i} \quad \mathrm{s.t.} \quad \vec{x}_{\mathrm{syn},i} \cong \vec{x}_{\mathrm{org},i} \quad \mathrm{and} \quad p_{g}\approx p_{o}
\end{equation}
\noindent where $p_{g}$ and $p_{o}$ represent the probability distribution of the generator and the original training signals, respectively. In order to derive a comprehensive $p_{g}$ for a given training set, the discriminator network should provide gradient information to $\mathcal{G}(\cdot)$ in every iteration. Toward this end, a metric should be employed to measure the discrepancy between the original and synthetic signals. The choice of such a metric (which is also known as an integral probability metric - IPM \cite{brezis2010functional}) is critical since it directly affects the learning curve of the generator network during training \cite{mroueh2017fisher}. Technically, an IPM measures the divergence between two probability distributions using a critic function $f(\cdot)$ as follows \cite{papoulis1990probability}:
\begin{equation}
    \sup_{f\in \mathcal{F}} \left [ \mathbb{E}_{\mathcal{G}(\mathbf{z}_{i}) \sim p_{g}} f(\mathcal{G}(\mathbf{z}_{i}))- \mathbb{E}_{\vec{x}_{\mathrm{org}} \sim p_{o}} f(\vec{x}_{\mathrm{org}}) \right ]
    \label{eq:discrepany}
\end{equation}
\noindent where $f(\cdot)$ is parametric and semidefinite \cite{sriperumbudur2012empirical}. Additionally, $\mathcal{F}$ is an independent function class defined in a non-Cartesian subspace. Basically, the interpretation of Eq.~\ref{eq:discrepany} in the context of training a GAN is finding a nonlinear critic function which asymptotically estimates the discrepancy between original and synthetic signals according to the constraints imposed by $\mathcal{F}$. Theoretically, there exist numerous function classes for Eq.~\ref{eq:discrepany} but only a limited number of those are practically integrable as an IPM into a generative model framework \cite{mroueh2017sobolev}. Some notable IPMs which have been successfully implemented for training different types of GANs are $\varphi$-divergence \cite{goodfellow2014generative} (used in the vanilla-GAN \cite{goodfellow2014generative}), Stein \cite{wang2016learning}, Wasserstein \cite{arjovsky2017towards} (for improving the stability and scalability of the generator networks), Cram\'{e}r \cite{bellemare2017cramer}, $\mu$-Fisher \cite{mroueh2017fisher}, and Sobolev \cite{mroueh2017sobolev}. Recently, the latter IPM has received more attention in the domain of speech synthesis using GANs since it is relatively more consistent with the latest architectures of the advanced speech-to-text transcription systems (e.g., DeepSpeech) \cite{esmaeilpour2022multi}. Such architectures employ a standard transformation layer known as the $\mathrm{mfcc}$ production which converts a single or multichannel input signal into a feature representation using Fourier transform. This transform is associated with the second degree function class defined in the Sobolev space ($\mathcal{F}_{S}$) \cite{mroueh2017sobolev}. The formal definition for $\mathcal{F}_{S}$ is as the following \cite{brezis2010functional}.
\begin{equation}
     \underbrace{\mathcal{S}^{\vartheta,2}(\mathcal{X})}_{\mathcal{F}_{S}}=\left\{f:\mathcal{X}\rightarrow\mathbb{R}^{d_{z}},\int_{\mathcal{X}}\left\|\nabla_{\mathbf{x}}f(\mathbf{x})\right\|^{2}\mu(\mathbf{x})d\mathbf{x} \right\} 
    \label{eq:sobolevSub}
\end{equation}
\noindent where $\mathcal{X}$ is an open subset of speech signal collection and $\vartheta$ indicates the degree of the function class. Additionally, $\mu(\cdot)\propto  p_{o}+p_{g}$ and it denotes a dominant probability density function \cite{brezis2010functional}. From the statistical and algebraic standpoints, $\mathcal{F}_{S}$ can be also defined as the following \cite{brezis2010functional,van1983matrix}:
\begin{equation}
     \mathcal{F}_{S} \cong  \sum \alpha_{\vartheta}\left | \tilde{f}(\mathbf{x}) \right |^{2} \quad \mathrm{and} \quad \mathrm{mfcc}(\vec{x}) \mapsto \mathbf{x}
    \label{eq:hilbertFour}
\end{equation}
\noindent where $\tilde{f}(\cdot)$ denotes a set of Fourier coefficients (i.e., frequency components) and $\alpha_{\vartheta}$ is set to be a nonzero scalar. Hence, due to the correlation between $\mathcal{F}_{S}$ and the $\mathrm{mfcc}$ feature vectors, substituting Eq.~\ref{eq:sobolevSub} into Eq.~\ref{eq:discrepany} most likely contributes to extract more informative gradients to the benefit of both $\mathcal{G}(\cdot)$ and $\mathcal{D}(\cdot)$ \cite{esmaeilpour2022multi}. The final outcome of this substitution is yielding the Sobolev-IPM (SIPM) with zero-boundary condition.

In practice, implementing the SIPM for training a GAN might be extremely challenging since to the best of our knowledge, there is no analytical upperbound limit for the integral term in Eq.~\ref{eq:sobolevSub} \cite{van1983matrix}. This increases the degree of nonlinearity for the critic function and results in delaying the convergence of the discriminator network during training \cite{esmaeilpour2022multi}. To address this issue, we propose a strict regularization term defined in the vector decomposition space \cite{van1983matrix} for $f(\cdot)$ partially aiming at offloading $\mathcal{D}(\cdot)$ from excessive computational overhead.

\textbf{Proposition:} Assuming $\lambda_{\nabla}$ and $\lambda_{f}$ represent the eigenvalue tensors for $\nabla f(\mathbf{x})$ and the critic function as defined in Eq.~\ref{eq:sobolevSub}, respectively. Then, we can find a non-zero parametric function $\Theta$ in such a way that:
\begin{equation}
    \min_{\lambda_{f}} \left | \lambda_{f} - \lambda_{\nabla} \right | \leq \eta \cdot \Theta\left \| \lambda_{f} \right \|_{F}
    \label{eq:reg2}
\end{equation}

\textbf{Proof:} Assuming $\eta \in \mathbb{R}_{+}^{d_{z}}$ and $\Theta^{-1}\left ( \lambda_{f} - \lambda_{\nabla}I  \right )\Theta$ is quasi-invertible. Then, according to the Bauer-Fike theorem \cite{van1983matrix} we can write:
\begin{equation}
     \sum \left | \lambda_{f} - \lambda_{\nabla} \right | \leqslant \left \| \lambda_{f} I- \Theta \right \|_{F}^{-1} \leqslant \eta \cdot \Theta\left \| \lambda_{f} \right \|_{F}
\end{equation}
\noindent where both $\lambda_{f}$ and $\Theta$ are in the closed-form. \hfill{$\square$} 

The intuition behind developing Eq.~\ref{eq:reg2} in an orthonormal Schur decomposition subspace is the possibility of constraining the critic function $f(\cdot)$ relative to an achievable upperbound (supremum). More specifically, our proposed regularizer binds the SIPM into a semidefinite region surrounded by this supremum value. Thus, we can have a tighter control over the nonlinearity of $f(\cdot)$ and train a more comprehensive yet stable GAN. However, the performance of our regularizer is highly dependent to accurately derive and maintaining $\Theta$. Toward this end, we define: 
\begin{equation}
   \Theta := \mathcal{FT}(\hat{\Theta}), \quad \hat{\Theta} \doteq \left \{{\hat{\theta}_{i,j}} \mid \hat{\theta}_{i,j} \sim \mathcal{N}(0,p_{\tau_{r}}) \right \}  
    \label{eq:Thetaaa}
\end{equation}
\noindent where $\mathcal{FT}(\cdot)$ represents the short-term Fourier transform \cite{griffin1984signal} for complying with the nature of $\mathrm{mfcc}$ features. Moreover, $p_{\tau_{r}}$ denotes the probability distribution of the simulated RIR filter sets \cite{yakura2018robust} as briefly mentioned in Section~\ref{sec:backg}. Employing such a distribution in modeling $\Theta$ enables us to incorporate the spectral features of the environmental settings into the regularization term of the critic function and potentially counteract with losing the quality of the speech signals upon running the defense GAN. In other words, substituting Eq.~\ref{eq:Thetaaa} into Eq.~\ref{eq:reg2} integrates some SM properties into $\mathcal{F}_{S}$ and forces the generator network to learn them. Eventually, this operation contributes to synthesize a naturally-sounding signal according to the characteristics of the environmental settings. In the following section, we provide a summary of our conducted experiments and analyze the performance of our proposed RSD-GAN relative to other defense approaches.

\begin{table*}[t]
\footnotesize
\centering
\scriptsize
\caption{Performance comparison of defense algorithms against a comprehensive set of targeted, non-targeted, white, and black-box adversarial signals. All the statistics are averaged over 10 times experiment repetitions. Outperforming results are in bold.}
\resizebox{\textwidth}{!}{\begin{tabular}{cccccccccccccccccccccc}
\hline
\multicolumn{1}{c||}{\multirow{2}{*}{Defense}} & \multicolumn{3}{c||}{Iteration ($\times 12,500$)}                                        & \multicolumn{3}{c||}{Modes ($\times 16.5$)}                                              & \multicolumn{3}{c||}{$\mathrm{GC}$ (per batch)}                                       & \multicolumn{3}{c||}{WER (\%)}                                                              & \multicolumn{3}{c||}{SLA (\%)}                                                              & \multicolumn{3}{c||}{segSNR}                                                                & \multicolumn{3}{c}{STOI}                                           \\ \cline{2-22} 
\multicolumn{1}{c||}{}                         & \multicolumn{1}{c|}{DPS}    & \multicolumn{1}{c|}{KLD}    & \multicolumn{1}{c||}{LGV}    & \multicolumn{1}{c|}{DPS}    & \multicolumn{1}{c|}{KLD}    & \multicolumn{1}{c||}{LGV}    & \multicolumn{1}{c|}{DPS}   & \multicolumn{1}{c|}{KLD}   & \multicolumn{1}{c||}{LGV}   & \multicolumn{1}{c|}{DPS}     & \multicolumn{1}{c|}{KLD}     & \multicolumn{1}{c||}{LGV}     & \multicolumn{1}{c|}{DPS}     & \multicolumn{1}{c|}{KLD}     & \multicolumn{1}{c||}{LGV}     & \multicolumn{1}{c|}{DPS}     & \multicolumn{1}{c|}{KLD}     & \multicolumn{1}{c||}{LGV}     & \multicolumn{1}{c|}{DPS}    & \multicolumn{1}{c|}{KLD}    & LGV    \\ \hline \hline
\multicolumn{1}{c||}{MRCA}                     & \multicolumn{1}{c|}{$-$}    & \multicolumn{1}{c|}{$-$}    & \multicolumn{1}{c||}{$-$}    & \multicolumn{1}{c|}{$-$}    & \multicolumn{1}{c|}{$-$}    & \multicolumn{1}{c||}{$-$}    & \multicolumn{1}{c|}{$-$}   & \multicolumn{1}{c|}{$-$}   & \multicolumn{1}{c||}{$-$}   & \multicolumn{1}{c|}{$29.01$} & \multicolumn{1}{c|}{$28.78$} & \multicolumn{1}{c||}{$25.31$} & \multicolumn{1}{c|}{$56.67$} & \multicolumn{1}{c|}{$58.42$} & \multicolumn{1}{c||}{$54.67$} & \multicolumn{1}{c|}{$16.09$} & \multicolumn{1}{c|}{$16.24$} & \multicolumn{1}{c||}{$17.88$} & \multicolumn{1}{c|}{$0.74$} & \multicolumn{1}{c|}{$0.75$} & $0.76$ \\ \hline
\multicolumn{1}{c||}{Dompteur}                 & \multicolumn{1}{c|}{$-$}    & \multicolumn{1}{c|}{$-$}    & \multicolumn{1}{c||}{$-$}    & \multicolumn{1}{c|}{$-$}    & \multicolumn{1}{c|}{$-$}    & \multicolumn{1}{c||}{$-$}    & \multicolumn{1}{c|}{$-$}   & \multicolumn{1}{c|}{$-$}   & \multicolumn{1}{c||}{$-$}   & \multicolumn{1}{c|}{$18.29$} & \multicolumn{1}{c|}{$16.25$} & \multicolumn{1}{c||}{$17.42$} & \multicolumn{1}{c|}{$61.91$} & \multicolumn{1}{c|}{$63.23$} & \multicolumn{1}{c||}{$59.70$} & \multicolumn{1}{c|}{$18.35$} & \multicolumn{1}{c|}{$18.16$} & \multicolumn{1}{c||}{$19.92$} & \multicolumn{1}{c|}{$0.77$} & \multicolumn{1}{c|}{$0.79$} & $0.79$ \\ \hline
\multicolumn{1}{c||}{A-GAN}                    & \multicolumn{1}{c|}{$0.71$} & \multicolumn{1}{c|}{$0.62$} & \multicolumn{1}{c||}{$0.57$} & \multicolumn{1}{c|}{$1.19$} & \multicolumn{1}{c|}{$1.14$} & \multicolumn{1}{c||}{$1.07$} & \multicolumn{1}{c|}{$619$} & \multicolumn{1}{c|}{$945$} & \multicolumn{1}{c||}{$877$} & \multicolumn{1}{c|}{$24.77$} & \multicolumn{1}{c|}{$21.62$} & \multicolumn{1}{c||}{$23.81$} & \multicolumn{1}{c|}{$65.19$} & \multicolumn{1}{c|}{$70.01$} & \multicolumn{1}{c||}{$61.16$} & \multicolumn{1}{c|}{$22.49$} & \multicolumn{1}{c|}{$19.72$} & \multicolumn{1}{c||}{$21.84$} & \multicolumn{1}{c|}{$0.81$} & \multicolumn{1}{c|}{$0.83$} & $0.78$ \\ \hline
\multicolumn{1}{c||}{CCD-GAN}                  & \multicolumn{1}{c|}{$1.76$} & \multicolumn{1}{c|}{$1.28$} & \multicolumn{1}{c||}{$1.42$} & \multicolumn{1}{c|}{$2.98$} & \multicolumn{1}{c|}{$1.49$} & \multicolumn{1}{c||}{$2.36$} & \multicolumn{1}{c|}{$433$} & \multicolumn{1}{c|}{$709$} & \multicolumn{1}{c||}{$612$} & \multicolumn{1}{c|}{$09.33$}  & \multicolumn{1}{c|}{$08.56$}  & \multicolumn{1}{c||}{$07.32$}  & \multicolumn{1}{c|}{$68.41$} & \multicolumn{1}{c|}{$70.89$} & \multicolumn{1}{c||}{$64.73$} & \multicolumn{1}{c|}{$22.76$} & \multicolumn{1}{c|}{$22.05$} & \multicolumn{1}{c||}{$24.39$} & \multicolumn{1}{c|}{$0.88$} & \multicolumn{1}{c|}{$0.89$} & $0.80$ \\ \hline
\multicolumn{1}{c||}{CD-GAN}                   & \multicolumn{1}{c|}{$1.87$} & \multicolumn{1}{c|}{$1.66$} & \multicolumn{1}{c||}{$1.59$} & \multicolumn{1}{c|}{$2.91$} & \multicolumn{1}{c|}{$2.77$} & \multicolumn{1}{c||}{$3.01$} & \multicolumn{1}{c|}{$512$} & \multicolumn{1}{c|}{$665$} & \multicolumn{1}{c||}{$564$} & \multicolumn{1}{c|}{$08.91$}  & \multicolumn{1}{c|}{$08.11$}  & \multicolumn{1}{c||}{$07.10$}  & \multicolumn{1}{c|}{$71.07$} & \multicolumn{1}{c|}{$73.52$} & \multicolumn{1}{c||}{$\mathbf{71.12}$} & \multicolumn{1}{c|}{$25.17$} & \multicolumn{1}{c|}{$24.78$} & \multicolumn{1}{c||}{$23.67$} & \multicolumn{1}{c|}{$0.87$} & \multicolumn{1}{c|}{$0.88$} & $0.81$ \\ \hline
\multicolumn{1}{c||}{SD-GAN}                   & \multicolumn{1}{c|}{$2.04$} & \multicolumn{1}{c|}{$1.79$} & \multicolumn{1}{c||}{$1.34$} & \multicolumn{1}{c|}{$3.85$} & \multicolumn{1}{c|}{$3.53$} & \multicolumn{1}{c||}{$3.29$} & \multicolumn{1}{c|}{$451$} & \multicolumn{1}{c|}{$562$} & \multicolumn{1}{c||}{$431$} & \multicolumn{1}{c|}{$08.22$}  & \multicolumn{1}{c|}{$\mathbf{07.53}$}  & \multicolumn{1}{c||}{$\mathbf{04.18}$}  & \multicolumn{1}{c|}{$71.39$} & \multicolumn{1}{c|}{$74.45$} & \multicolumn{1}{c||}{$69.97$} & \multicolumn{1}{c|}{$26.34$} & \multicolumn{1}{c|}{$25.22$} & \multicolumn{1}{c||}{$25.51$} & \multicolumn{1}{c|}{$0.89$} & \multicolumn{1}{c|}{$0.90$} & $0.82$ \\ \hline
\multicolumn{1}{c||}{RSD-GAN}                  & \multicolumn{1}{c|}{$\mathbf{2.89}$} & \multicolumn{1}{c|}{$\mathbf{2.23}$} & \multicolumn{1}{c||}{$\mathbf{2.96}$} & \multicolumn{1}{c|}{$\mathbf{4.18}$} & \multicolumn{1}{c|}{$\mathbf{4.77}$} & \multicolumn{1}{c||}{$\mathbf{4.15}$} & \multicolumn{1}{c|}{$\mathbf{211}$} & \multicolumn{1}{c|}{$\mathbf{171}$} & \multicolumn{1}{c||}{$\mathbf{189}$} & \multicolumn{1}{c|}{$\mathbf{06.11}$}  & \multicolumn{1}{c|}{$08.08$}  & \multicolumn{1}{c||}{$04.91$}  & \multicolumn{1}{c|}{$\mathbf{75.88}$} & \multicolumn{1}{c|}{$\mathbf{76.60}$} & \multicolumn{1}{c||}{$70.51$} & \multicolumn{1}{c|}{$\mathbf{34.12}$} & \multicolumn{1}{c|}{$\mathbf{35.68}$} & \multicolumn{1}{c||}{$\mathbf{34.81}$} & \multicolumn{1}{c|}{$\mathbf{0.96}$} & \multicolumn{1}{c|}{$\mathbf{0.97}$} & $\mathbf{0.93}$ \\ \hline
\multicolumn{22}{c}{{Average computational complexity comparison for defense algorithms (in metric seconds) per ensemble batch size of 512 during training \cite{che2016mode}: A-GAN: 1.17, CCD-GAN: 1.02, CD-GAN: 1.06, SD-GAN:1.03, and RSD-GAN: 0.72.}} \\
\multicolumn{22}{c}{\textit{Recall:} DPS: DeepSpeech $\mid$ KLD: Kaldi $\mid$ LGV: Lingvo $\mid$ Modes: this is averaged over all the batches \cite{che2016mode} $\mid$ $\mathrm{GC}$: this is computed for the discriminator network with static batch size \cite{che2016mode}.} 

\end{tabular}}
\label{table:defenseTech}
\vspace{-15pt}
\end{table*}

\section{Experiments}
\label{sec:exp}
Our benchmarking datasets for training the RSD-GAN are the Mozilla Common Voice (MCV \cite{MozillaCommonVoiceDataset}) and LibriSpeech \cite{panayotov2015librispeech} which both include above thousands hours of speech signals distributed over diverse utterances. Following a very common practice in the domain of speech adversarial attack and defense experimentation \cite{carlini2018audio, qin2019imperceptible, yakura2018robust, esmaeilpour2021cyclic}, we also run all the algorithms on a portion of the aforementioned datasets. Therefore, we randomly select 35,000 samples separately from MCV and LibriSpeech collections. We dedicate 30\% of such signals for crafting $\vec{x}_{\mathrm{adv}}$ using different attack algorithms and keep the rest for training the defense approaches.

The architecture which we design for our RSD-GAN is somewhat similar to the CCD-GAN, however without employing the class-conditional platform. For the generator network, we implement four consecutive residual blocks with $4\times 4 \times 16$ channels followed by two $16 \rightarrow 4 \rightarrow 1$ non-local layers. All these hidden layers are accompanied by weight normalization, orthogonal initialization \cite{SaxeMG13}, and $\tanh$ activation function. For the discriminator network, we implement two stacked residual blocks with $4\times 4 \times 2$ channels and two $3\times 3 \times 1$ convolutional layers. Finally, there is one linear logit layer ($\rightarrow 1$) prior to a softmax function. This network employs the SIPM with our proposed regularization term (i.e., Eq.~\ref{eq:reg2}) for measuring the discrepancy between $p_{g}$ and $p_{o}$. Toward deriving a comprehensive $\Theta$ for Eq.~\ref{eq:Thetaaa}, we follow the protocol explained in \cite{kallinger2006multi, yakura2018robust} and use their predefined filter sets.

For targeted attacks, namely C\&W, Yakura's, the robust attack, Imperio, Metamorph, and the CIPM attack, we randomly assign 15 incorrect phrases to $\hat{\mathbf{y}}_{i}$ (with various lengths) and run their associated optimization algorithms (e.g., Eq.~\ref{eq:generalformulation}). Upon the convergence of these attacks, we achieve 15 adversarial signals for every given $\vec{x}_{\mathrm{org}}$. Since we cannot manually assign such incorrect phrases to non-targeted and black-box attacks (e.g., the multi-objective optimization attack (MOOA) \cite{khare2018adversarial} and the genetic algorithm attack (GAA) \cite{taori2019targeted}), we run each of these algorithms 15 times to craft a collection of different adversarial signals.

In order to evaluate the performance of the defense algorithms against the aforementioned adversarial signals, we measure three categories of statistical metrics. The first category includes three estimators for the accountability of the GANs. More specifically, we compute the total number of iterations for the generator network prior to collapse, the number of learned modes in every batch with the static size of 512, and the rounds of required gradient computations ($\mathrm{GC}$) according to the pipeline characterized in \cite{che2016mode}. Technically, larger values for the first two estimators interpret as the higher stability of the generative model in runtime. However, $\mathrm{GC}$ indicates the computational load on the discriminator network and ideally it should be minimized. 

The second category contains the conventional yet effective word error rate (WER) and sentence-level accuracy (SLA) metrics as defined in the following \cite{qin2019imperceptible,esmaeilpour2022multi}:
\begin{equation}
   \mathrm{WER} = (\hat{S}+\hat{I}+\hat{D})/n_{\mathbf{\hat{y}}_{i}}\times 100 \mid \mathrm{SLA} = n_{\mathrm{ct}}/n_{\mathrm{tas}}\times 100
    \label{eq:WERSLA}
\end{equation}
\noindent where $\hat{S}$, $\hat{I}$, and $\hat{D}$ denote the number of phrase substitution, insertion, and deletion, respectively. Moreover, $n_{\mathbf{\hat{y}}_{i}}$ indicates the total number of predefined adversarial phrases. Regarding the SLA's mechanism, $n_{\mathrm{ct}}$ and $n_{\mathrm{tas}}$ stand for the number of correctly transcribed signals and the total number of adversarial signals crafted for each victim model (e.g. DeepSpeech), respectively. Finally, the third category accommodates rigorous signal quality metrics, namely segmental signal-to-noise ratio (segSNR) \cite{baby2019sergan} and the short-term objective intelligibility (STOI) \cite{taal2011algorithm}. We employ such estimators to assess the level of distortions imposed on the signals upon passing them through the defense algorithms. By definition, both segSNR and STOI metrics yield higher values for more reliable defense algorithms. Table~\ref{table:defenseTech} summarizes our achieved results on the above-mentioned evaluation categories. As shown, our RSD-GAN dominantly outperforms other defense algorithms in terms of generative model accountability. This implies that not only our proposed regularizer (i.e., Eq.~\ref{eq:reg2}) can significantly increase the number of learned modes for $\mathcal{G}(\cdot)$ and keep the model stable up to above 20,000 iterations, but also it remarkably decreases the number of gradient computations for $\mathcal{D}(\cdot)$. Additionally, for almost half of the cases the RSD-GAN competitively surpasses other dense approaches with reference to the WER and SLA metrics. Finally, the achieved results for the signal quality metrics corroborates the effectiveness of our novel RIR encoding technique (i.e., Eq.~\ref{eq:Thetaaa}) into the SIPM formulation. In other words, the RSD-GAN makes a better trade-off between protecting speech transcription systems against adversarial attacks and preserving the quality of the synthetic signals.


\section{Conclusion}
In this paper, we introduced a novel defense approach for protecting some advanced speech-to-text transcription systems. Our algorithm is based on synthesizing a new signal acoustically very similar to the given malicious input aiming at bypassing the potential adversarial perturbation. Toward this end, we developed a non-conditional GAN architecture and imposed a solid regularization on the discriminator network for improving accuracy, maintaining stability, and reducing the computational complexity of the entire model. However, we noticed that the performance of our RSD-GAN considerably drops for really long multispeaker signals. We are strictly determined to address this issue in our future works. 

\balance

\bibliographystyle{IEEEtran}
\bibliography{IEEEabrv,mybib}

\begin{thebibliography}{10}
\providecommand{\url}[1]{#1}
\csname url@samestyle\endcsname
\providecommand{\newblock}{\relax}
\providecommand{\bibinfo}[2]{#2}
\providecommand{\BIBentrySTDinterwordspacing}{\spaceskip=0pt\relax}
\providecommand{\BIBentryALTinterwordstretchfactor}{4}
\providecommand{\BIBentryALTinterwordspacing}{\spaceskip=\fontdimen2\font plus
\BIBentryALTinterwordstretchfactor\fontdimen3\font minus
  \fontdimen4\font\relax}
\providecommand{\BIBforeignlanguage}[2]{{%
\expandafter\ifx\csname l@#1\endcsname\relax
\typeout{** WARNING: IEEEtran.bst: No hyphenation pattern has been}%
\typeout{** loaded for the language `#1'. Using the pattern for}%
\typeout{** the default language instead.}%
\else
\language=\csname l@#1\endcsname
\fi
#2}}
\providecommand{\BIBdecl}{\relax}
\BIBdecl

\bibitem{vila2018end}
L.~C. Vila, C.~Escolano, J.~A. Fonollosa, and M.~R. Costa-Jussa, ``End-to-end
  speech translation with the transformer.'' in \emph{IberSPEECH}, 2018, pp.
  60--63.

\bibitem{MozillaImplementation}
M.~Implementation, ``Mozilla speech recognition project: Deepspeech,''
  https://github.com/mozilla/DeepSpeech, 2017.

\bibitem{povey2011kaldi}
D.~Povey, A.~Ghoshal, G.~Boulianne, L.~Burget, O.~Glembek, N.~Goel,
  M.~Hannemann, P.~Motlicek, Y.~Qian, P.~Schwarz \emph{et~al.}, ``The kaldi
  speech recognition toolkit,'' in \emph{IEEE Works Autom Speech Recog
  Underst}, 2011.

\bibitem{shen2019lingvo}
J.~Shen, P.~Nguyen, Y.~Wu, Z.~Chen, M.~X. Chen, Y.~Jia, A.~Kannan, T.~Sainath,
  Y.~Cao, C.-C. Chiu \emph{et~al.}, ``Lingvo: a modular and scalable framework
  for sequence-to-sequence modeling,'' \emph{arXiv preprint arXiv:1902.08295},
  2019.

\bibitem{schonherr2020imperio}
L.~Sch{\"o}nherr, T.~Eisenhofer, S.~Zeiler, T.~Holz, and D.~Kolossa, ``Imperio:
  Robust over-the-air adversarial examples for automatic speech recognition
  systems,'' in \emph{Annual Comp Secur Appl Conf}, 2020, pp. 843--855.

\bibitem{carlini2018audio}
N.~Carlini and D.~Wagner, ``Audio adversarial examples: Targeted attacks on
  speech-to-text,'' in \emph{IEEE Secur Privacy Workss}, 2018, pp. 1--7.

\bibitem{wang2020towards}
Q.~Wang, B.~Zheng, Q.~Li, C.~Shen, and Z.~Ba, ``Towards query-efficient
  adversarial attacks against automatic speech recognition systems,''
  \emph{IEEE Transactions on Information Forensics and Security}, vol.~16, pp.
  896--908, 2020.

\bibitem{papernot2018characterizing}
N.~Papernot, ``Characterizing the limits and defenses of machine learning in
  adversarial settings,'' \emph{PhD Thesis presented to The Pennsylvania State
  University}, 2018.

\bibitem{esmaeilpour2020sound}
\BIBentryALTinterwordspacing
M.~Esmaeilpour, P.~Cardinal, and A.~L. Koerich, ``From environmental sound
  representation to robustness of 2d cnn models against adversarial attacks,''
  \emph{Applied Acoustics}, vol. 195, p. 108817, 2022. [Online]. Available:
  \url{https://www.sciencedirect.com/science/article/pii/S0003682X22001918}
\BIBentrySTDinterwordspacing

\bibitem{esmaeilpour2022multi}
------, ``Multi-discriminator sobolev defense-gan against adversarial attacks
  for end-to-end speech systems,'' \emph{IEEE Transactions on Information
  Forensics and Security}, vol.~17, pp. 2044--2058, 2022.

\bibitem{goodfellow2014generative}
I.~Goodfellow, J.~Pouget-Abadie, M.~Mirza, B.~Xu, D.~Warde-Farley, S.~Ozair,
  A.~Courville, and Y.~Bengio, ``Generative adversarial nets,'' in \emph{Adv
  Neural Inf Process Syst}, 2014, pp. 2672--2680.

\bibitem{samangouei2018defensegan}
P.~Samangouei, M.~Kabkab, and R.~Chellappa, ``Defense-{GAN}: Protecting
  classifiers against adversarial attacks using generative models,'' in
  \emph{Intl Conf Learn Repres}, 2018.

\bibitem{van1983matrix}
C.~F. Van~Loan and G.~H. Golub, \emph{Matrix computations}.\hskip 1em plus
  0.5em minus 0.4em\relax Johns Hopkins University Press, 1983.

\bibitem{mroueh2017sobolev}
Y.~Mroueh, C.~Li, T.~Sercu, A.~Raj, and Y.~Cheng, ``Sobolev {GAN},'' in
  \emph{6th Intl Conf Learn Repres}, 2018.

\bibitem{akhtar2018threat}
N.~Akhtar and A.~Mian, ``Threat of adversarial attacks on deep learning in
  computer vision: A survey,'' \emph{Ieee Access}, vol.~6, pp.
  14\,410--14\,430, 2018.

\bibitem{yakura2018robust}
H.~Yakura and J.~Sakuma, ``Robust audio adversarial example for a physical
  attack,'' in \emph{28th Intl J Conf Artif Intell}, 2018, pp. 5334--5341.

\bibitem{habets2006room}
E.~A. Habets, ``Room impulse response generator,'' \emph{Technische
  Universiteit Eindhoven, Tech. Rep}, vol.~2, no. 2.4, p.~1, 2006.

\bibitem{davis1980comparison}
S.~Davis and P.~Mermelstein, ``Comparison of parametric representations for
  monosyllabic word recognition in continuously spoken sentences,'' \emph{IEEE
  Trans Acoust, Speech, Signal Process}, vol.~28, no.~4, pp. 357--366, 1980.

\bibitem{qin2019imperceptible}
Y.~Qin, N.~Carlini, G.~Cottrell, I.~Goodfellow, and C.~Raffel, ``Imperceptible,
  robust, and targeted adversarial examples for automatic speech recognition,''
  in \emph{Intl Conf Mach Learn}, 2019, pp. 5231--5240.

\bibitem{chen2020metamorph}
T.~Chen, L.~Shangguan, Z.~Li, and K.~Jamieson, ``Metamorph: Injecting inaudible
  commands into over-the-air voice controlled systems,'' in \emph{Netw Distrib
  Syst Secur Symp}, 2020.

\bibitem{esmaeilpour2022towards}
M.~Esmaeilpour, P.~Cardinal, and A.~L. Koerich, ``Towards robust speech-to-text
  adversarial attack,'' in \emph{ICASSP 2022-2022 IEEE International Conference
  on Acoustics, Speech and Signal Processing (ICASSP)}.\hskip 1em plus 0.5em
  minus 0.4em\relax IEEE, 2022, pp. 2869--2873.

\bibitem{hu2019adversarial}
S.~Hu, X.~Shang, Z.~Qin, M.~Li, Q.~Wang, and C.~Wang, ``Adversarial examples
  for automatic speech recognition: Attacks and countermeasures,'' \emph{IEEE
  Communications Magazine}, vol.~57, no.~10, pp. 120--126, 2019.

\bibitem{esmaeilpour2021towards}
M.~Esmaeilpour, ``Towards reliable data-driven sound recognition models:
  developing attack and defense algorithms,'' Ph.D. dissertation,
  D\'{e}partement de G\'{e}nie Logiciel et des TI, {\'E}cole de technologie
  sup{\'e}rieure ({\'E}TS), Montr\'{e}al, Qu\'{e}bec, Canada, 2021.

\bibitem{das2018adagio}
N.~Das, M.~Shanbhogue, S.-T. Chen, L.~Chen, M.~E. Kounavis, and D.~H. Chau,
  ``Adagio: Interactive experimentation with adversarial attack and defense for
  audio,'' \emph{arXiv preprint arXiv:1805.11852}, 2018.

\bibitem{esmaeilpour2020class}
M.~Esmaeilpour, P.~Cardinal, and A.~L. Koerich, ``Class-conditional defense gan
  against end-to-end speech attacks,'' in \emph{IEEE Intl Conf Acoust, Speech
  and Signal Process}, 2021, pp. 2565--2569.

\bibitem{eisenhofer2021dompteur}
T.~Eisenhofer, L.~Sch{\"o}nherr, J.~Frank, L.~Speckemeier, D.~Kolossa, and
  T.~Holz, ``Dompteur: Taming audio adversarial examples,'' in \emph{30th
  USENIX Secur Symp}, 2021, pp. 2309--2326.

\bibitem{latif2018adversarial}
S.~Latif, R.~Rana, and J.~Qadir, ``Adversarial machine learning and speech
  emotion recognition: Utilizing generative adversarial networks for
  robustness,'' \emph{arXiv preprint arXiv:1811.11402}, 2018.

\bibitem{esmaeilpour2021cyclic}
M.~Esmaeilpour, P.~Cardinal, and A.~L. Koerich, ``Cyclic defense gan against
  speech adversarial attacks,'' \emph{IEEE Signal Processing Letters}, vol.~28,
  pp. 1769--1773, 2021.

\bibitem{mcfee2015librosa}
B.~McFee, C.~Raffel, D.~Liang, D.~P. Ellis, M.~McVicar, E.~Battenberg, and
  O.~Nieto, ``Librosa: Audio and music signal analysis in python,'' in
  \emph{14th Python in Science Conf}, vol.~8, 2015.

\bibitem{esmaeilpour2020detection}
M.~Esmaeilpour, P.~Cardinal, and A.~L. Koerich, ``Detection of adversarial
  attacks and characterization of adversarial subspace,'' in \emph{IEEE Intl
  Conf Acoust, Speech and Signal Process}, 2020, pp. 3097--3101.

\bibitem{hosseini2018multi}
\BIBentryALTinterwordspacing
E.~Hosseini{-}Asl, Y.~Zhou, C.~Xiong, and R.~Socher, ``A multi-discriminator
  cyclegan for unsupervised non-parallel speech domain adaptation,'' in
  \emph{Interspeech 2018, 19th Annual Conference of the International Speech
  Communication Association, Hyderabad, India, 2-6 September 2018}.\hskip 1em
  plus 0.5em minus 0.4em\relax {ISCA}, 2018, pp. 3758--3762. [Online].
  Available: \url{https://doi.org/10.21437/Interspeech.2018-1535}
\BIBentrySTDinterwordspacing

\bibitem{freeman1970efficient}
J.~Freeman and D.~Ford, ``The efficient calculation of the cartesian geometry
  of non-cartesian structures,'' \emph{WIT Transactions on Engineering
  Sciences}, vol.~16, 1970.

\bibitem{brezis2010functional}
H.~Brezis, \emph{Functional analysis, Sobolev spaces and partial differential
  equations}.\hskip 1em plus 0.5em minus 0.4em\relax Springer Science \&
  Business Media, 2010.

\bibitem{mroueh2017fisher}
Y.~Mroueh and T.~Sercu, ``Fisher {GAN},'' in \emph{Adv in Neural Inf Proc Sys
  30: Annual Conf on Neural Inf Proc Sys}, 2017, pp. 2513--2523.

\bibitem{papoulis1990probability}
A.~Papoulis, \emph{Probability and statistics}.\hskip 1em plus 0.5em minus
  0.4em\relax Prentice-Hall, Inc., 1990.

\bibitem{sriperumbudur2012empirical}
B.~K. Sriperumbudur, K.~Fukumizu, A.~Gretton, B.~Sch{\"o}lkopf, G.~R. Lanckriet
  \emph{et~al.}, ``On the empirical estimation of integral probability
  metrics,'' \emph{Electronic Journal of Statistics}, vol.~6, pp. 1550--1599,
  2012.

\bibitem{wang2016learning}
Y.~Feng, D.~Wang, and Q.~Liu, ``Learning to draw samples with amortized stein
  variational gradient descent,'' in \emph{33rd Conf Uncert Artif Intell},
  2017.

\bibitem{arjovsky2017towards}
M.~Arjovsky and L.~Bottou, ``Towards principled methods for training generative
  adversarial networks,'' in \emph{5th Intl Conf Learn Repres}, 2017.

\bibitem{bellemare2017cramer}
M.~G. Bellemare, I.~Danihelka, W.~Dabney, S.~Mohamed, B.~Lakshminarayanan,
  S.~Hoyer, and R.~Munos, ``The cramer distance as a solution to biased
  wasserstein gradients,'' \emph{CoRR}, vol. abs/1705.10743, 2017.

\bibitem{griffin1984signal}
D.~Griffin and J.~Lim, ``Signal estimation from modified short-time fourier
  transform,'' \emph{IEEE Trans Acoust, Speech, Signal Process}, vol.~32,
  no.~2, pp. 236--243, 1984.

\bibitem{che2016mode}
T.~Che, Y.~Li, A.~P. Jacob, Y.~Bengio, and W.~Li, ``Mode regularized generative
  adversarial networks,'' in \emph{5th Intl Conf on Learning Representations,
  {ICLR}, Toulon, France}, 2017.

\bibitem{MozillaCommonVoiceDataset}
M.~commonvoice.mozilla.org, ``Mozilla common voice dataset,''
  https://voice.mozilla.org/en/datasets, 2019.

\bibitem{panayotov2015librispeech}
V.~Panayotov, G.~Chen, D.~Povey, and S.~Khudanpur, ``Librispeech: an asr corpus
  based on public domain audio books,'' in \emph{IEEE Intl Conf Acoust, Speech
  and Signal Process}, 2015, pp. 5206--5210.

\bibitem{SaxeMG13}
A.~M. Saxe, J.~L. McClelland, and S.~Ganguli, ``Exact solutions to the
  nonlinear dynamics of learning in deep linear neural networks,'' in \emph{2nd
  Intl Conf Learn Repres}, 2014.

\bibitem{kallinger2006multi}
M.~Kallinger and A.~Mertins, ``Multi-channel room impulse response shaping-a
  study,'' in \emph{2006 IEEE International Conference on Acoustics Speech and
  Signal Processing Proceedings}, vol.~5.\hskip 1em plus 0.5em minus
  0.4em\relax IEEE, 2006, pp. V--V.

\bibitem{khare2018adversarial}
S.~Khare, R.~Aralikatte, and S.~Mani, ``Adversarial black-box attacks on
  automatic speech recognition systems using multi-objective evolutionary
  optimization,'' \emph{arXiv preprint arXiv:1811.01312}, 2018.

\bibitem{taori2019targeted}
R.~Taori, A.~Kamsetty, B.~Chu, and N.~Vemuri, ``Targeted adversarial examples
  for black box audio systems,'' in \emph{IEEE Security and Privacy Works},
  2019, pp. 15--20.

\bibitem{baby2019sergan}
D.~Baby and S.~Verhulst, ``Sergan: Speech enhancement using relativistic
  generative adversarial networks with gradient penalty,'' in \emph{IEEE Intl
  Conf Acoust, Speech, Signal Process}, 2019, pp. 106--110.

\bibitem{taal2011algorithm}
C.~H. Taal, R.~C. Hendriks, R.~Heusdens, and J.~Jensen, ``An algorithm for
  intelligibility prediction of time--frequency weighted noisy speech,''
  \emph{IEEE Trans Audio, Speech, Lang Process}, vol.~19, no.~7, pp.
  2125--2136, 2011.

\end{thebibliography}

\end{document}